\documentclass{rnc6}
\usepackage[latin1]{inputenc}
\usepackage{aeguill}
\usepackage{comment}
\usepackage{xspace}

\makeatletter

\def\url{%
  \begingroup
  \catcode`\~=\active
  \catcode`:=12
  \def~{\raisebox{-.8ex}{\textasciitilde}}%
  \@url}
\def\@url#1{\leavevmode
  \hskip 0pt plus 50pt \penalty 1000 \hskip 0pt plus -50pt
  \mbox#1\endgroup}

\def\footurl{%
  \begingroup
  \catcode`\~=\active
  \catcode`:=12
  \def~{\raisebox{-.8ex}{\textasciitilde}}%
  \@footurl}
\def\@footurl#1{\footnote{#1}\endgroup}

\makeatother

\newcommand{\rnd}{\ensuremath{\diamond}\xspace}
\newcommand{\err}{\ensuremath{\varepsilon}\xspace}

\begin{document}
\sloppy
\binoppenalty=10000
\relpenalty=10000
\doublehyphendemerits 40000


\begin{frontmatter}

\title{The Generic Multiple-Precision Floating-Point Addition
With Exact Rounding (as in the MPFR Library)}

\author{Vincent Lefèvre}

\ead{Vincent.Lefevre@loria.fr}
\ead[url]{http://www.vinc17.org/research/}

\address{INRIA Lorraine, 615 rue du Jardin Botanique,
54602 Villers-lès-Nancy Cedex, France}

\begin{abstract}
We study the multiple-precision addition of two positive floating-point
numbers in base $2$, with exact rounding, as specified in the MPFR
library, i.e. where each number has its own precision. We show how
the best possible complexity (up to a constant factor that depends
on the implementation) can be obtain.
\end{abstract}

\begin{keyword}
multiple precision \sep floating point \sep addition \sep exact rounding
\end{keyword}

\end{frontmatter}


\section{Introduction}
\label{intro}

In this paper, we consider the multiple-precision floating-point
addition with exact rounding, as specified in the MPFR
library\footurl{http://www.mpfr.org/}: the inputs are two (binary)
floating-point numbers $x$ and $y$ of precision $m \geq 2$ and
$n \geq 2$, a target precision $p \geq 2$ and a rounding mode \rnd,
and the output is $\rnd(x + y)$, i.e. the exact value $x + y$ rounded
to the target precision in the given rounding mode, and a ternary value
giving the sign of $\rnd(x + y) - (x + y)$.

By ``addition'', we mean here the ``true'' addition, that is $x + y$
where $x$ and $y$ have the same sign, and $x - y$ where $x$ and $y$
have opposite signs. For the sake of simplicity, we restrict to the
addition of positive values for $x$ and $y$ in the following of the
paper. In fact, this is how it is implemented in MPFR: indeed, the
addition and subtraction functions call an auxiliary function, ignoring
the signs of the input numbers (they are regarded as positive).

The addition seems to be a very simple function to implement, as
being a basic function, easy to understand. This is unfortunately
not true, in particular under the MPFR specifications (where the
inputs and the output may have different precisions and the output
must be the exactly rounded result), because many different cases
need to be considered, and it is easy to forget one, both in the
implementation of the addition and in the tests. For a long time,
the MPFR addition had been buggy (more precisely, some rare special
cases were not handled correctly) and inefficient (also in some
rare special cases, for which the time complexity was exponential).
I completely rewrote the addition in October 2001. The presentation
given in this paper is more or less based on the same ideas as the
ones used in the new MPFR implementation, with some non-theoretical
differences, mainly due to some MPFR internals.

First, the floating-point system is introduced in Section~\ref{fp}.
The main computation steps (from which the algorithm can be
deduced) and the complexity are presented in Section~\ref{comp}.
Section~\ref{impl} deals with the MPFR implementation. We finally
conclude in Section~\ref{concl}.


\section{The Floating-Point System}
\label{fp}

\subsection{The Floating-Point Representation}
\label{fp-repr}

We consider a floating-point system in base $2$. The results presented
in this paper can naturally be extended to other fixed even bases, but
we choose the base $2$ (this is the base of the floating-point system
in MPFR) to make the notations easier to understand.

In our system, an object may contain a special value, like NaN (not
a number) or an infinity, zero (possibly signed, as in the IEEE-754
standard\cite{IEEE754} and in MPFR), or an non-zero real number that
can be written:
\[s \times 0.b_1 b_2 b_3 \ldots b_p \times 2^e,\]
where $s = \pm 1$ is the \emph{sign}, the $b_i$'s are binary digits
($0$ or $1$) forming the \emph{mantissa}, $e$ is the \emph{exponent}
(a bounded integer\footnote{In MPFR, the exponent is between $1 - 2^{30}$
and $2^{30} - 1$.}), and $p$ is an integer greater or equal to $1$,
called the \emph{precision}. In MPFR, the precision $p$ is not fixed;
it is attached to each object. Non-zero real values are normalized,
i.e.\ $b_1 \ne 0$, that is in base $2$, $b_1 = 1$. The system does not
have subnormals (i.e. numbers with $b_1 = 0$) as they are not really
useful with a huge exponent range (like in MPFR) and would make the
algorithms and the code much more complex.

In the following of the paper, we consider only positive input numbers
(as said in the introduction), i.e. numbers that can be written:
$0.1 b_2 b_3 \ldots b_p \times 2^e$.

\subsection{Rounding}
\label{fp-round}

When adding two positive numbers $x$ and $y$ of respective precisions
$m$ and $n$, the result is not necessarily representable in the target
precision $p$; it must be rounded, according to one of the rounding
modes chosen by the user, similar to the IEEE-754 rounding modes:

\begin{itemize}

\item rounding to minus infinity (downwards): we return the largest
floating-point number in precision $p$ that is less or equal to $x + y$;

\item rounding to plus infinity (upwards): we return the smallest
floating-point number in precision $p$ that is greater or equal to $x + y$;

\item rounding towards zero: we round downwards, since $x + y > 0$;

\item rounding to the nearest: we return the floating-point number
in precision $p$ that is the closest to $x + y$. Halfway cases are
specified by the implementation; if $p \geq 2$ (this is required
by MPFR), then we can choose the round-to-even rule, like in the
IEEE-754 standard and in MPFR: we return the only number that has
an even mantissa, i.e. with $b_p = 0$.

\end{itemize}

Note that the returned result must be the rounding of the exact result;
this requirement is called \emph{correct} or \emph{exact rounding}.

In addition to the rounded result, a ternary value is returned, giving
the sign of $\rnd(x + y) - (x + y)$, where \rnd denotes the chosen
rounding mode: a positive number means that the rounded result is
greater or equal to the exact result, a negative number means that
the rounded result is less or equal to the exact result, and $0$
means that the returned result is the exact result.

Moreover, it is possible that the exponent of the rounded result is
not in the exponent range, in which case an overflow is generated.
This case does lot lead to any practical or theoretical difficulty
and is beyond the scope of this paper.

How the exact result of a canonical infinite mantissa $0.1 b_2 b_3 \ldots$
(where the number of zero bits is infinite) is rounded can be expressed
as a function of the bit $r = b_{p+1}$ following the truncated $p$-bit
mantissa, called the \emph{rounding bit}, and $s = b_{p+2} \vee b_{p+3}
\vee \ldots$, called the \emph{sticky bit}, as summarized in
Table~\ref{fig:rs} (we recall that the result is positive).

\begin{table}[htbp]
\begin{center}
\begin{tabular}{|c|c|c|c|}
\hline
$r$ / $s$ & downwards & upwards & to the nearest \\
\hline
$0$ / $0$ &   exact   &  exact  &     exact      \\
\hline
$0$ / $1$ &    $-$    &   $+$   &      $-$       \\
\hline
$1$ / $0$ &    $-$    &   $+$   &   $-$ / $+$    \\
\hline
$1$ / $1$ &    $-$    &   $+$   &      $+$       \\
\hline
\end{tabular}
\end{center}
\caption{A $-$ in the table means that the mantissa of the exactly
rounded result is $0.1 b_2 b_3 \ldots b_p$, i.e. the truncated
exact mantissa. A $+$ in the table means that one needs to add
$2^{-p}$ to the truncated mantissa (leading to an exponent change
if all the $b_i$'s up to $b_p$ are $1$). The $-$~/~$+$ corresponds
to the halfway cases, and the round-to-even rule is applied, that
is: $-$ if $b_p = 0$, $+$ if $b_p = 1$.}
\label{fig:rs}
\end{table}

Note: We did not mention the ternary value, as it can easily be deduced
from Table~\ref{fig:rs} (telling how the mantissa is rounded). Also,
like the rounding modes towards $-\infty$ and towards $+\infty$, the
returned ternary value needs to be negated if the result is negative
(not considered in this paper).


\section{The Main Computation Steps and the Complexity}
\label{comp}

We still denote the precisions of the input numbers $x$ and $y$ and
the result by $m$, $n$ and $p$ respectively.

The addition of two positive floating-point numbers $x$ and $y$ of
respective exponents $e_x$ and $e_y$ consists in:
\begin{enumerate}
\item ordering $x$ and $y$ so that $e_x \geq e_y$,
\item computing the exponent difference $d = e_x - e_y$,
\item shifting the mantissa of $y$ by $d$ positions to the right,
\item adding the mantissa of $x$ and the shifted mantissa of $y$
and rounding the result (shifting it by $1$ position to the right
if there is a carry),
\item computing the exponent of the result: $e_x$ or $e_x + 1$ if
there is a carry.
\end{enumerate}

This method is very inefficient if many trailing bits of $x$ or $y$
(possibly all the bits of $y$) do not have any influence on the
result, for instance:
\[0.101010000010010001 + 0.10001 \times 2^{-9}\]
rounded to $4$ bits. The exactly rounded result and the ternary value
can be deduced from only the first $6$ bits $101010$ of $x$ (and none
for $y$), knowing the fact that its first mantissa bit is always $1$.

So, we are interested in taking into account as few input bits as
possible (the possible \emph{hole} between the least significant bit
of $x$ and the most significant bit of $y$ must also be detected). We
do not have any particular knowledge about the input numbers $x$ and
$y$ (and the result); we assume that the mantissa bits are $0$ and $1$
with equal probabilities after some given position and that $x$ and
$y$ are independent numbers. Of course, this is not necessarily a good
assumption, but this will be discussed when it has an importance.

The addition can be written $x + y = t + \err$, where $t$ is the
\emph{main term}, computed with the first $p+2$ bits of $x$ and
the corresponding $\max(p+2-d,0)$ bits of $y$, and \err is the
\emph{error term}, satisfying $0 \leq \err < 2^{e_x - p - 1}$.
This can graphically be represented by:
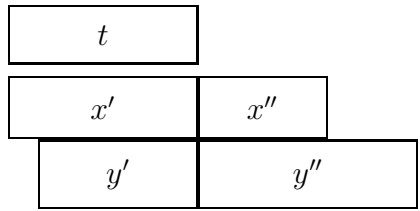
\[
\begin{array}{r@{}l}
\fbox{\begin{minipage}{20mm}\centerline{$t$}\end{minipage}} & \\[2mm]
\fbox{\begin{minipage}{20mm}\centerline{$x'$}\end{minipage}} &
\fbox{\begin{minipage}{12mm}\centerline{$x''$}\end{minipage}} \\[1mm]
\fbox{\begin{minipage}{16mm}\centerline{$y'$}\end{minipage}} &
\fbox{\begin{minipage}{24mm}\centerline{$y''$}\end{minipage}} \\
\end{array}
\]
where $x''$ may be empty and either $y'$ or $y''$ may be empty.

The \emph{main term} $t$ is computed and written in time $\Theta(p)$;
indeed, an $\Omega(p)$ time is necessary to fill the $p+2$ bits, and
a linear
time is obviously sufficient. There are many ways to deal with all the
different cases (the mantissas of $x$ and $y$ may completely overlap,
partially overlap in numerous ways, or even not overlap at all, and
some parts of the result may need to be filled with zeros); a carry
detection can also be performed by looking at the most significant
bits of $x$ and $y$ first. More will be said in Section~\ref{impl},
about the implementation in MPFR. However this is not an important
point here, as long as the complexity is in $\Theta(p)$.

The \emph{error term} allows to obtain the truncated mantissa, the
rounding bit and the sticky bit (Section~\ref{fp-round}). First, if
the computation of the main term has lead to a carry, then $p+3$ bits
of the result have really been computed. This case can be regarded
as if there were no carry and the first iteration of the processing
described below were already performed (then, this is only a matter
of implementation). So, for the sake of simplicity, let us consider
that $p+2$ bits of the result have been computed, let $u$ denote the
\emph{weight}\footnote{This is the corresponding power of $2$; for
instance, the weight of the bit $1$ in $0.001$ is $2^{-3}$.} of the
bit $p+2$ (so, $0 \leq \err < 2u$), and let $f$ denote
its value ($0$ or $1$)\footnote{If a carry was generated, consider
only the first $p+2$ bits of the result in $t$, and the bit $p+3$ is
taken into account in \err.}, that we call the \emph{following} bit.
Table~\ref{fig:fe} gives the rounding bit $r$ and the sticky bit $s$
as a function of the following bit $f$ and the error $\err$.

\begin{table}[htbp]
\begin{center}
\begin{tabular}{|c|c||c|c|}
\hline
$f$ &   $\err$   & $r$ & $s$ \\
\hline
$0$ & $\err = 0$ & $=$ & $0$ \\
\hline
$0$ & $\err > 0$ & $=$ & $1$ \\
\hline
$1$ & $\err < u$ & $=$ & $1$ \\
\hline
$1$ & $\err = u$ & $+$ & $0$ \\
\hline
$1$ & $\err > u$ & $+$ & $1$ \\
\hline
\end{tabular}
\end{center}
\caption{For $r$, an $=$ means that the rounding bit is the bit $p+1$
of the temporary result $t$, and a $+$ means that $1$ must be added to
the bit $p+1$ of $t$ (and the carry must propagate).}
\label{fig:fe}
\end{table}

Combining Tables \ref{fig:rs} and \ref{fig:fe}, we get Table~\ref{fig:rfe}.
Now we may need to determine the sign of $\err - fu$ (depending on the
cases given by Table~\ref{fig:rfe}). This is done with an iteration over
the remaining bits of $x$ and $y$.

\begin{table}[htbp]
\begin{center}
\begin{tabular}{|c|c|c||c|c||c|c|c|}
\hline
$r_t$ & $f$ &   $\err$   & $r$ & $s$ & downwards & upwards & to the nearest \\
\hline
 $0$  & $0$ & $\err = 0$ & $0$ & $0$ &   exact   &  exact  &     exact      \\
\hline
 $0$  & $0$ & $\err > 0$ & $0$ & $1$ &    $-$    &   $+$   &      $-$       \\
\hline
 $0$  & $1$ & $\err < u$ & $0$ & $1$ &    $-$    &   $+$   &      $-$       \\
\hline
 $0$  & $1$ & $\err = u$ & $1$ & $0$ &    $-$    &   $+$   &   $-$ / $+$    \\
\hline
 $0$  & $1$ & $\err > u$ & $1$ & $1$ &    $-$    &   $+$   &      $+$       \\
\hline
 $1$  & $0$ & $\err = 0$ & $1$ & $0$ &    $-$    &   $+$   &   $-$ / $+$    \\
\hline
 $1$  & $0$ & $\err > 0$ & $1$ & $1$ &    $-$    &   $+$   &      $+$       \\
\hline
 $1$  & $1$ & $\err < u$ & $1$ & $1$ &    $-$    &   $+$   &      $+$       \\
\hline
 $1$  & $1$ & $\err = u$ & $0$ & $0$ &   exact   &  exact  &     exact      \\
\hline
 $1$  & $1$ & $\err > u$ & $0$ & $1$ &    $-$    &   $+$   &      $-$       \\
\hline
\end{tabular}
\end{center}
\caption{The first three columns give all the possible cases for the
rounding bit of the main term, the following bit $f$ and the error $\err$.
The next two columns give the corresponding values of the rounding bit $r$
and the sticky bit $s$ (once the error has been taken into account).
The last three columns give information for the rounded result and the
ternary value; in the last two cases (lines), a carry is added to the
mantissa before the rounding (and this may lead to an exponent change,
but has no effect on how the rounding is performed --- implementations
must just take care that the ulp is different when rounding upwards).}
\label{fig:rfe}
\end{table}

\begin{itemize}

\item If $f = 0$, we need to distinguish the cases $\err = 0$ and $\err > 0$.
We have: $\err > 0$ if and only if at least a trailing bit (of $x$ or $y$)
is $1$. In particular, if $y < u$, then the most significant bit of $y$
(always $1$) is a trailing bit (said otherwise, $\err \geq y$); so, in this
case, $\err > 0$. Otherwise, one needs to test the trailing bits, the one
after the other until a $1$ is found, and in the worst case ($\err = 0$),
all the trailing bits need to be tested. As a consequence, the worst-case
complexity in the case $f = 0$ is in $\Theta(m+n+p)$.

Is there a best order to test the trailing bits? Under the condition
that we do not have any particular knowledge on the input numbers,
there is no best order\footnote{One can argue that the real numbers
naturally are logarithmically distributed, so that the probability
to have a $0$ at position $i$ is higher than the one to have a $1$,
and the difference decreases as a function of $i$ \cite{FelGoo1976a}.
Therefore, in a very theoretical point of view, if the time of each
test is seen as a constant, it would be better to start by the least
significant bits! Of course, since the probabilities get very close
to $1/2$ very quickly, one would not see any difference in practice.}.
One can choose one of the following two possibilities, for instance:
\begin{itemize}
\item Test the trailing bits of $x$, then the trailing bits of $y$ (or
the other way round) until a non-zero bit is found.
\item Test trailing bits of both $x$ and $y$ at the same time. This may
be an interesting choice as some numbers tend to have an exact mantissa
with few non-zero digits (like small integers), thus many trailing zeros.
Testing trailing bits of $x$ and $y$ concurrently may allow to avoid
such difficult cases.
\end{itemize}

\item If $f = 1$, we need to distinguish the cases $\err < u$, $\err = u$
and $\err > u$. Let $d$ denote the exponent difference so that the bits
$x_i$ and $y_{i-d}$ have the same weight ($d$ is the shift count to align
the mantissas). Let $q$ be the first integer such that the trailing bits
$x_q$ and $y_{q-d}$ are equal (when a bit is not represented, it is $0$).
If these bits are $0$'s, then $\err < u$. Otherwise (i.e. if these bits
are $1$'s), $\err \geq u$. The equality $\err = u$ can be decided as in
the case $f = 0$ ($\err > u$ if and only if at least one the untested
bits is $1$).

The best way is to start with bit $p+3$ and loop over the
increasing positions, until $q$ is found (and more if one has to decide
if $\err = u$). If the $x$ mantissa or the $y$ mantissa has entirely
been read and $q$ has not been found yet, then it is not necessary to
go further, as we can deduce that $\err < u$; in other words, there
is no possible carry with bits from only one mantissa.

Up to the position $q-1$, we have $x_i + y_{i-d} = 1$, as one of the bits
is $0$ and the other is $1$. When $x$ and $y$ overlap, it is necessary to
test at least one bit (as one of these two bits has an importance in the
sign of $\err - u$). Concerning the other bits, the result can be deduced
from only one test (like in the case $f = 0$), but an oracle would be
needed, and it is not possible to do better without any particular
knowledge. Here is an example: let us considered $x = 0.101111100101$
($12$-bit precision), $y = 0.11010 \times 2^{-7}$ ($5$-bit precision),
and a $2$-bit target precision. The mantissas are aligned in the following
way:
\[
\begin{array}{cr}
  & 0.101111100101 \\
+ &        0.11010 \\
\end{array}
\]
Though on this particular example, testing the last five bits $0$ is
sufficient to deduce that the exact result is less than $0.11$, all the
bits are tested from the left to the right. Moreover, if either $x$ or $y$
(but not both) had more bits, e.g.\ $y = 0.11010111001 \times 2^{-7}$,
then testing these bits would not be necessary as they cannot generate
a carry to reach $0.11$; however, if $y = 0.110110000 \times 2^{-7}$,
testing the following four bits would be necessary to deduce that the
result is $0.11$ exactly.

\end{itemize}

In the case $f = 1$, the loop is performed over the increasing positions
from the bit $p+3$, grabbing the bit of $x$ and the bit of $y$ having the
same weight. The same loop can be performed in the case $f = 0$, though
this is not the only solution as said above. Of course, special cases
must be taken into account: the $x$ mantissa does not necessarily overlap
with the aligned $y$ mantissa (as the most significant bit of $y$ may
come after some trailing bits of $x$, some trailing bits of $x$ may come
after the least significant bit of $y$, some trailing bits of $y$ may
come after the least significant bit of $x$), and the most significant
bit of $y$ may come after the least significant bit of $x$ (\emph{hole}
between the $x$ mantissa and the $y$ mantissa), any \emph{missing} bit
being regarded as $0$. At each iteration, the mantissa of the temporary
result has the form: $0.1 z_2 z_3 \ldots z_p r f f f \ldots f f f$ with
an error in the interval $[0,2)$ ulp\footnote{Unit in the Last Place:
here, the weight of the last bit $f$ of the temporary result.}. One
iterates as long as the bits after the (temporary) rounding bit are
identical. This basically corresponds to the Table Maker's Dilemma,
that occurs to exactly round any function (see \cite{LefMulTis1998b},
for instance).

The time complexity is in $\Omega(p)$ and in $O(m+n+p)$. In the worst
case, it is in $\Theta(m+n+p)$. In average (if the bits are $0$ and
$1$ with equal probabilities and input numbers are independent), the
complexity is in $\Theta(p)$, as the probability to need to test $k$
trailing bits decreases exponentially (in $2^{-k}$).


\section{The MPFR Implementation}
\label{impl}

In this section, we present the implementation of the addition in MPFR.
To keep the paper from being too technical, we do not give many details
(and the reader can read the source code, as MPFR is distributed under
the GNU Lesser General Public License). A complete proof would also be
very hard to read and check (unless it could be mechanically checked);
so, such a proof is not provided.

Let us start with the representation of MPFR numbers. Non-special MPFR
numbers have a sign (accessed with C macros), a mantissa, an exponent
(some C integer) and a precision (also some C integer). The mantissa is
represented by an array of \emph{limbs}; a limb is an unsigned integer
(having 32 bits or 64 bits, depending on the C implementation), as defined
in the GMP library\footurl{http://www.swox.com/gmp/}, on which MPFR is
based. All the bits of a limb are used to represent bits of the mantissa
in the conventional binary representation\footnote{GMP now allows to use
some bits for the carries, called \emph{nail bits}. They are not supported
yet in MPFR. One should note that contrary to integer operations, redundancy
provided by nail bits would probably not be very interesting here due to
the discontinuity of the rounding function.}.
The mantissa is normalized, i.e. its most significant bit is always $1$.
Since the precision is not necessarily a multiple of the limb size, some
bits of the lowest mantissa limb are not significant and are always $0$
(except in temporary values).

The computation steps presented in Section~\ref{comp} were bit based
(as this is more regular and easier to understand for a theoretical
analysis). But working on single bits in a software implementation
would not be very efficient. Base operations must be performed by
blocks; some limbs may still need to be split into two parts as the
$y$ mantissa must be aligned with the $x$ mantissa.

In addition to the particular cases that arise in the bit-based case,
we need to distinguish the case where the exponent difference is a
multiple of the limb size and the other case, needing the $y$ mantissa
to be shifted (this is usually done on the fly). We also need to take
into account all the cases related to the block boundaries (for
instance, where the rounding bit lies in a limb).

First, the main term is computed, but there are differences with the
bit-based version. As the array holding the target mantissa does not
necessarily have the room for $p+2$ bits and we want to avoid an
inefficient memory allocation for $p+2$ bits and copy, the temporary
rounding bit and the following bit are stored in C integer variables
\texttt{rb} and \texttt{fb} (determined on the fly, as soon as they
are known); in this way, then can also be handled more efficiently.
The second difference is all the bits of the target array
are used for this computation (in fact, this is more or less necessary,
as the low-level GMP functions do not perform any masking).

For the main term, we want to add the most significant parts $x'$ and
$y'$. If $y$ does not overlap with the main term ($y'$ is empty), we
just copy $x'$ to the target array and zero the least significant limbs
of the target if the target has a greater precision than $x$. Now, let
us assume that $y$ overlaps with the main term. With GMP, we cannot
shift and add with a single operation; therefore these operations have
to be performed separately. First, with a GMP function, we copy the
most significant part of $y$, shifted if need be, to the target array
and we zero the limbs of the target that have not been touched: the
most and/or least significant limbs, if the exponent and the precision
of $y$ are small enough. Then, with another GMP function, we add the
most significant part of $x$ to the target. If a carry is generated,
we increment the exponent (unless we already had the maximum exponent,
in which case we generate an overflow) and shift the result to the right;
a bit is
lost due to the shift but it is either the rounding bit or a following
bit, and if necessary, the following bits are tested and the rounding
can be performed if they are not all equal.

Then, the non-significant bits of the target are taken into account;
this occurs only if:
\begin{itemize}
\item the rounding bit is still unknown (otherwise these bits have
already been taken into account before the shift due to the carry,
as said above), and
\item there are non-significant bits.
\end{itemize}

At this time, the rounding bit and the following bit may still be
unknown; in this case, they will be determined as soon as possible
from the trailing parts $x''$ and $y''$. The loops are performed
on the increasing positions (by blocks), as mentioned at the end
of Section~\ref{comp}; moreover the cases $f = 0$ and $f = 1$ are
considered together (and not separated as in Section~\ref{comp}).

The iterations depend on the current status concerning $x$ and $y$.
Here are the different cases that may arise during the iterations:
\begin{itemize}
\item $x''$ has not entirely been read and $y''$ has not been read yet.
\item $x''$ and $y''$ overlap.
\item $x''$ has not entirely been read and $y''$ has entirely been read.
\item $x''$ has entirely been read and $y''$ has not been read yet.
\item $x''$ has entirely been read and $y''$ has not entirely been read.
\item $x''$ and $y''$ have entirely been read.
\end{itemize}

In the overlapping case, at each iteration, a limb of $x$ and the
corresponding limb of $y$ (built from two different limbs if $y$ must
be shifted) are added. The possible carry is taken into account, and
the loop ends as soon as the result is $0$ (all its bits are $0$) for
$f = 0$ or the maximum limb value \verb|MP_LIMB_T_MAX| (all its bits
are $1$) for $f = 1$.

We have focused on the differences coming from the computations by
blocks. The whole details may be found in the MPFR code.


\section{Conclusion}
\label{concl}

We have presented the generic multiple-precision floating-point addition
with exact rounding, as specified in the MPFR library, first in a rather
theoretical point of view, then considering the current implementation
in MPFR. The theoretical analysis could give a more regular description
of the implementation, by ignoring the fact that bits are grouped
into words in a computer memory. It could help to improve the current
implementation (the fact that the cases $f = 0$ and $f = 1$ are
considered together is probably not a very good idea, though it
reduces the risk of forgetting particular cases).

The subtraction could be dealt with in a similar manner, in future
work. This is a bit more complicated due to a possible cancellation
(when subtracting very close numbers).

Full mechanically-checked proofs could also be considered, using the
theoretical analysis to define the main notions.



\end{document}